\documentclass[a4paper,11pt]{article}
\usepackage{graphicx}
\usepackage{bm}
\usepackage{amsmath}
\usepackage{enumerate}
\usepackage{color}
\usepackage{braket}
\usepackage{amsmath}
\usepackage{enumerate}

\renewcommand{\vec}[1]{\mathbf{#1}}

\newcommand{\idxModel}{{\cal M}}

\begin{document}
\title{
Complex sequencing rules of birdsong can be explained by simple hidden Markov processes} 
\date{\today}

\author{
Kentaro Katahira$^{1,2,3}$, Kenta Suzuki$^{3,4}$, Kazuo Okanoya$^{1,3}$ and Masato Okada$^{1,2,3}$ \\
$^{1}$Japan Science Technology Agency, ERATO, Okanoya Emotional Information Project, \\
2-1 Hirosawa, Wako, 351-0198 Japan \\
$^{2}$Graduate School of Frontier Sciences, \\
The University of Tokyo, Chiba 277-5861, Japan \\
$^{3}$RIKEN Brain Science Institute 2-1 Hirosawa, Wako, 351-0198 Japan \\
$^{4}$Graduate School of Science and Engineering, \\ 
Saitama University, Saitama 338-0825, Japan
}

\maketitle

\begin{abstract}
Complex sequencing rules observed in birdsongs provide an opportunity to investigate the neural mechanism for generating complex sequential behaviors. To relate the findings from studying birdsongs to other sequential behaviors, it is crucial to characterize the statistical properties of the sequencing rules in birdsongs. However, the properties of the sequencing rules in birdsongs have not yet been fully addressed. In this study, we investigate the statistical propertiesof the complex birdsong of the Bengalese finch ({\it Lonchura striata var. domestica}). Based on manual-annotated syllable sequences, we first show that there are significant higher-order context dependencies in Bengalese finch songs, that is, which syllable appears next depends on more than one previous syllable. This property is shared with other complex sequential behaviors. We then analyze acoustic features of the song and show that higher-order context dependencies can be explained using first-order hidden state transition dynamics with redundant hidden states. This model corresponds to hidden Markov models (HMMs), well known statistical models with a large range of application for time series modeling. The song annotation with these models with first-order hidden state dynamics agreed well with manual annotation, the score was comparable to that of a second-order HMM, and surpassed the zeroth-order model (the Gaussian mixture model (GMM)), which does not use context information. Our results imply that the hierarchical representation with hidden state dynamics may underlie the neural implementation for generating complex sequences with higher-order dependencies. 
\end{abstract}

\section{Introduction}
Humans can generate complex sequential behaviors such as speech and musical performance. These sequences are typically composed of sequences of actions with complex sequencing rules. How our brain generates such complex sequences is difficult to understand in a straightforward manner since underlying neural circuits are complex and it is difficult to precisely explore neural circuits in human or primate brains. A solution to this issue may be given by studying songbirds~\cite{doupe1999birdsong}. In particular, Bengalese finches sing with apparently more complex sequencing rules with branching points ~\cite{honda1999acoustical,okanoya2006bengalese,okanoya1998adult}, than does the zebra finch, whose songs are composed of a stereotyped syllable sequence and extensively used for birdsong studies. Bengalese finch songs have been receiving attention as a model of variable sequential behavior, from neurophysiological~\cite{hosino2000lesion,okanoya2004song,sakata2006real,sakata2009social} and theoretical~\cite{katahira2007neural, yamashita2008developmental,jin2009generating} view points. 

To understand the mechanism for the variable sequences of the Bengalese finch song, it is important to characterize the song sequences from a statistical view point. However, the statistical properties of the Bengalese finch song have not been extensively studied. We first demonstrate that the Bengalese finch song has higher-order context dependency: each syllable appears depending on the most recent as well as more than one recent syllable. (For example in Figure~\ref{fig:G1_1automaton}A, the emission probability of syllable ``c'' and ``d'' depends not only on the adjacent syllable ``b'' but also on the preceding syllables ``a'' and ``c''. ) This property has been mentioned in previous studies~\cite{okanoya2006bengalese}. However, we demonstrated its statistical significance for the first time. 

We then investigated the statistical mechanism for explaining higher-order dependencies observed in Bengalese finch songs. To do this, we used the Bayesian inference method and a model selection technique. We applied hidden Markov models (HMMs) with various context dependencies to the acoustic features of a Bengalese finch song and selected a suitable model based on the Bayesian model comparison, its predictive performance, and the degree of agreement with manual annotation. As a result, we found that the first-order HMM, in which the current state appears depending only on the last state, is sufficient and suitable for describing the Bengalese finch song. Perhaps this is a counterintuitive result since the song sequences have higher-order dependency as we mentioned. This is due to a mechanism by which the first-order HMM can generate apparently complex sequences, which we describe in this paper. These results imply that the songbird brain has parsimonious neural representation for generating apparently complex sequences. Also, these results support the branching-chain mechanism, which has been proposed in theoretical studies~\cite{katahira2007neural,jin2009generating}, for generating Bengalese finch song sequences. 

\section{Results}
\begin{figure}[tbph]
\begin{center} \includegraphics[width=\linewidth]
{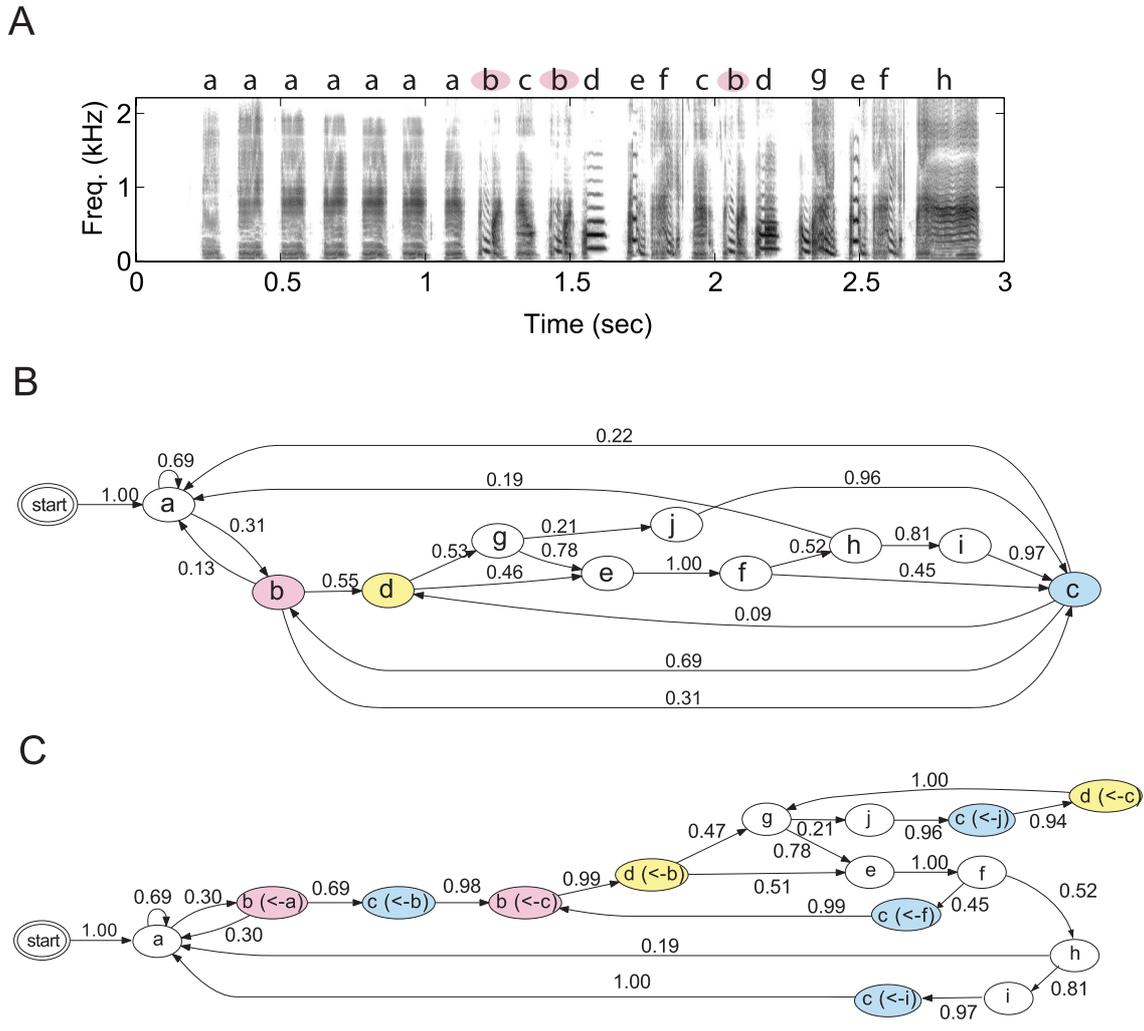} 
\end{center} 
\caption{Example of sonogram of Bengalese finch song and its syllable label sequence. (A) Sonogram of Bengalese finch (BF09) with syllable labels annotated by three human experts. Labeling was done based on visual inspection of sonogram and syllables with similar spectrogram given same syllable. (B) Bigram automaton representation (transition diagram) of syllable sequences obtained from same song set as (A). Ellipses represent one syllable and arrows with values represent transitional probabilities. Rare transitions with probabilities $<0.01$ are omitted. (C) POMM representation of same sequences as (B). Syllables that have significant higher-order dependency on preceding syllables (colored states in (B)) are divided into distinct states depending on preceding syllables (context).} 
\label{fig:G1_1automaton} 
\end{figure}

We analyzed the songs of 16 normal adult male Bengalese finches (See Method for details.) An example of the sonogram (sound spectrogram) of a Bengalese finch song is shown in Figure~\ref{fig:G1_1automaton}A. The Bengalese finch song consists of acoustically continuous segments, called ``song elements'' or ``syllables'' (in this paper, we used the term ``syllable'') which are separated by silent intervals. Bengalese finch songs are often analyzed by assigning a label to acoustically similar syllables, usually based on visual inspection on the sonogram. Following this approach, we first analyzed the statistical properties of the syllable label strings. We then directly analyzed the acoustic features using statistical models and compared the results to those of an analysis on manual annotated labels. 

\paragraph{Higher-order context dependency in syllable sequences of Bengalese finch song}

We show that the song syllable strings annotated by three human experts in analysis of birdsong have higher-order context dependency. The three experts labeled based on visual inspection on sound spectrographs. We cross checked by computing Fleiss's $\kappa$ coefficient~\cite{fleiss1971measuring}, which measures the degree of agreement among more than two annotators (see Methods). As a result, the $\kappa$-coefficients were 0.972 $\pm$ 0.028 (mean $\pm$ s.d.) for the 16 birds, and all within the range of ``Almost perfect agreement'', indicating annotation by the three experts was reliable. Hereafter, we use the labeling results by only one of the labeling experts. 

We conducted a hypothesis test for each syllable to verify whether the preceding syllables of the syllable being tested affects the occurrence probability of the next syllable (see Methods). We found more than one significant second-order dependency in all 16 birds. When we restricted the analysis to non-repeated syllables, significant syllables were found in the songs of 11 birds. In total, there were 33 significant syllable triplets (21 for non-repetitive syllables) of 72 candidate syllables. An example is shown in Figure~\ref{fig:G1_1automaton}. In this song, the syllables labeled ``b'' are preceded by either ``a'' or ``c'', and are followed by ``a'', ``c'', or ``d'' (Fig.~\ref{fig:G1_1automaton}B). If syllable ``c'' precedes syllable ``b'', the transition probability from ``b'' to ``d'' is 0.99, but if we do not care about the preceding syllable of "b", the transition probabilities to syllables ``a'', ``d'', and ``c'' are 0.13, 0.55 and 0.31, respectively. There was a significant difference between these two probability distributions ($\chi^2(2) = 511.9898, p < 10^{-5}$), indicating that preceding syllables ``a'' and ``c'' had a significant effect on the transition probabilities from syllable ``b''. 

This second-order context-dependency can be visually captured by splitting the syllables into distinct states depending on the preceding states. Such representation, in which different states are allowed to emit the same syllable, is regarded as a model called the partially observable Markov model (POMM)~\cite{jin2009generating}, thus we call this the POMM representation. For example, the state corresponding to ``b'' in Fig.~\ref{fig:G1_1automaton}B is divided into states $b(\leftarrow a)$ and $b(\leftarrow c)$ depending on the preceding syllables (a or c). From the first-order HMM (Fig.~\ref{fig:G1_1automaton}B), it may seem that transition from syllable ``b'' to ``a'', ``d'', and ``c'' are random, but with the POMM representation (Fig.~\ref{fig:G1_1automaton}C), we can capture the tendency that if ``a'' precedes, ``b'' is followed by the syllables ``a'' or ``c'', but if ``c'' precedes, ``b'' is followed by the syllable ``d'' almost deterministically. In addition, from the POMM representation, we can see that the syllable sequences ``bcbd'' and ``jcd'' are sung in chunks. Taken together, we conclude that the sequencing rules of the Bengalese finch song have higher order Markov dependency, and cannot be described using a simple Markov process, where states and syllables have one-to-one mapping. Nevertheless, a transition diagram in which a simple Markov process is implicitly assumed has been often used for analyzing Bengalese finch songs because of its simplicity~\cite{sakata2006real,wohlgemuth2010linked}. We need to be careful when interpreting such representation if we derive the information about variability of the syllable sequence. Even if branching points are found in the diagram, it does not necessarily imply that the following syllable is variable (or stochastic): it may be a stereotyped given more than one previous syllables. 

\paragraph{Hidden Markov model analysis on acoustic features} 
Next, we searched for a suitable statistical description of the Bengalese finch song directly from acoustic feature data extracted from the audio-signal of the song. We used HMMs~\cite{rabiner1989tutorial}, which have been widely used for time-series data modeling, including human speech recognition and also birdsong annotation (but used differently from the present study)~\cite{kogan1998automated}. In HMMs, the observed data are assumed to be generated from probabilistic distributions (here, we use a single Gaussian distribution) associated with hidden states, which are usually assumed to be generated from a first-order Markov process. We extend the HMMs to incorporate second-order transition dynamics of hidden states, in accordance with the above results (see Methods section). In addition, we also include the ``zeroth-order HMM'', which has the same structure as the first-order HMM but without a transition matrix (i.e., context dependency). This model exactly corresponds to the GMM, a statistical model used for data clustering. HMMs with higher than second-order are in principle possible to construct. However, because of their computational cost, which  increases exponentially with order, we did not examine them. Furthermore, as can be expected from the following results, such higher-order HMMs would not produce better results than with the first-order HMM. In addition to the order in hidden Markov processes, there is a degree of freedom in the model structure, that is, the number of hidden states, denoted by $K$. Based on the Bayesian model selection technique (see Methods section) and cross-validation, we explored the best model for describing the Bengalese finch song within a set of our models. 

\paragraph{Model comparison} 

\begin{figure}[tbph]
\begin{center}
\includegraphics[width=0.95\linewidth]{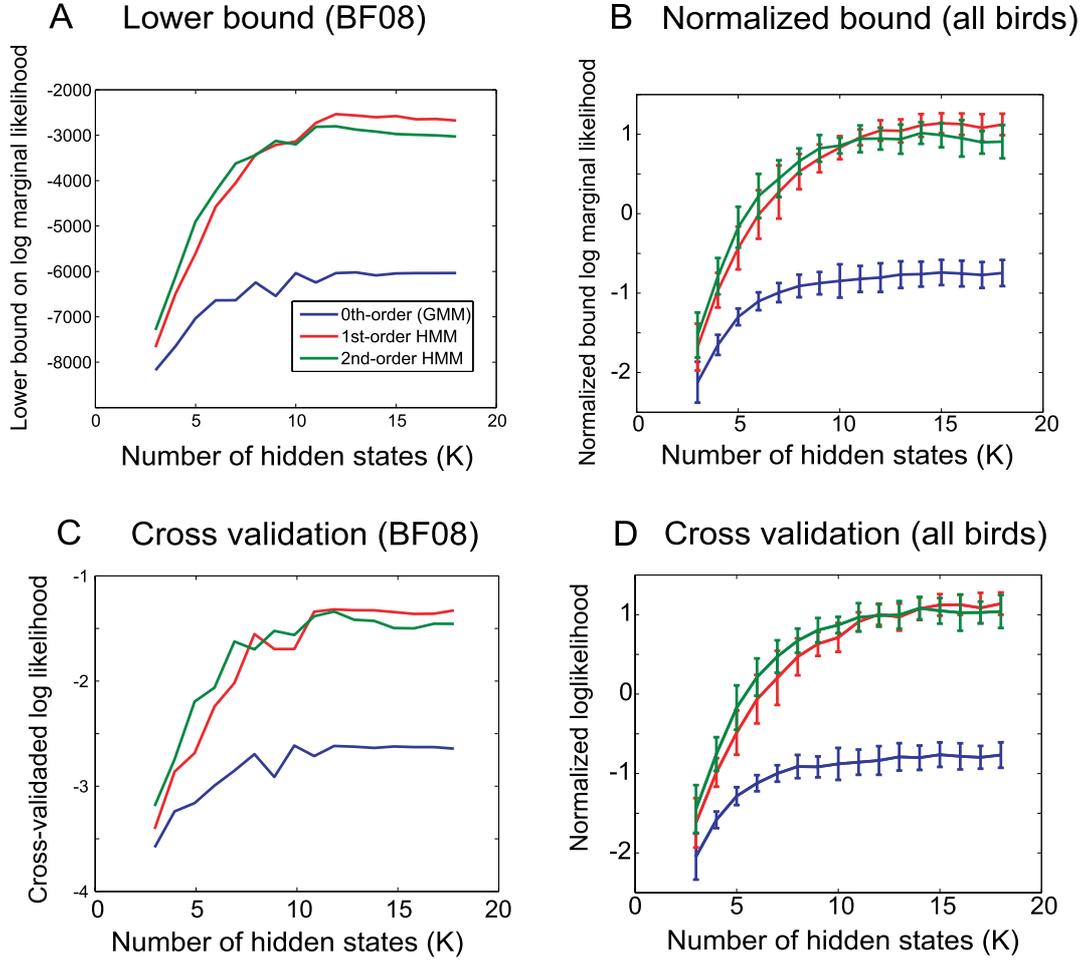} 
\end{center}
\caption{
Comparison of statistical models with various states $K$ and model 
orders on acoustic features of Bengalese finch song. (A-B) Plot of lower 
bound on marginal log likelihood. Larger this bound, the more 
appropriate model is for representing given data. For both cases, 
first-order HMM gave largest bound provided there was sufficient number 
of states available. (C-D) Cross validated log-likelihood on test data 
sets obtained from same bird on same date but ten different bouts from 
those used for training model. (A,C): representative bird (BF08). (B, D): 
average over all birds on normalized value. Error bars indicate standard 
deviation.
} 
\label{fig:freeEnergy_birdsong}
\end{figure}

Figures~\ref{fig:freeEnergy_birdsong}A and B compare the lower bound on the log marginal likelihood, which is a model selection criterion (see Methods section), among various hidden states ($K$) and orders of Markov processes. The model that gives the largest lower bound is regarded as the most appropriate for the given data. This criterion automatically embodies a {\it Bayesian Occam's razor}~\cite{mackay1992bayesian,bishop2006pattern}: a model with many parameters are given a larger penalty than one with fewer parameters. Thus, the simplest model, which can sufficiently describe the given data set, is selected. We see that for both representative birds (Figure~\ref{fig:freeEnergy_birdsong}A) and the average over all birds (Figure~\ref{fig:freeEnergy_birdsong}B), the second-order HMMs showed a larger bound when a small number of states were given ($K$). However, for a large number of states, the first-order HMMs gave the largest bound. Similar results ($n=13/16$) were obtained for almost all the song data from other birds we analyzed. The exceptions ($n=3$) were for the songs in which no significant second-order context dependency (excluding the repetitive syllables) was observed. These results with the approximate Bayesian model comparison suggest that the first-order HMM would suffice when sufficient hidden-states are available. We interpret these results in more detail in the Discussion section. 

To evaluate how well the models describe the statistics of song acoustic features more directly, we computed the predictive performance of the models based on cross-validated log-likelihood on test data (that were not used for model training). The test data consisting of ten bouts for each bird were constructed from the song recorded from the same bird on the same date with the training data. The results for the two birds are shown in Figures~\ref{fig:freeEnergy_birdsong}C and D. Similar tendencies with the lower bound were observed: with smaller $K$, the second-order HMM provided the highest likelihood, and after increasing $K$, the first-order HMM had the best performance or was comparable to the second-order HMM. The number of states and model order showd significantly affected the predictive performance (two-way ANOVA, $p < 10^{-5}$ and $p < 10^{-5}$, respectively). The first- and second-order HMMs performed better than the zeroth-order model (Tukey's multiple comparison, $p < 0.05$), and there was significant difference between the first- and second-order models ($p < 0.05$). Also, for the selected models based on the lower bound for each model order, the first- and second-order HMMs performed better than the zeroth-order model ($p < 0.05$) and the first- and HMMs performed better than the second-order model  ($p < 0.05$). 

\paragraph{Comparison with manual annotation}
\begin{figure}[tbhp]
\begin{center}
\includegraphics[width=.75\linewidth]{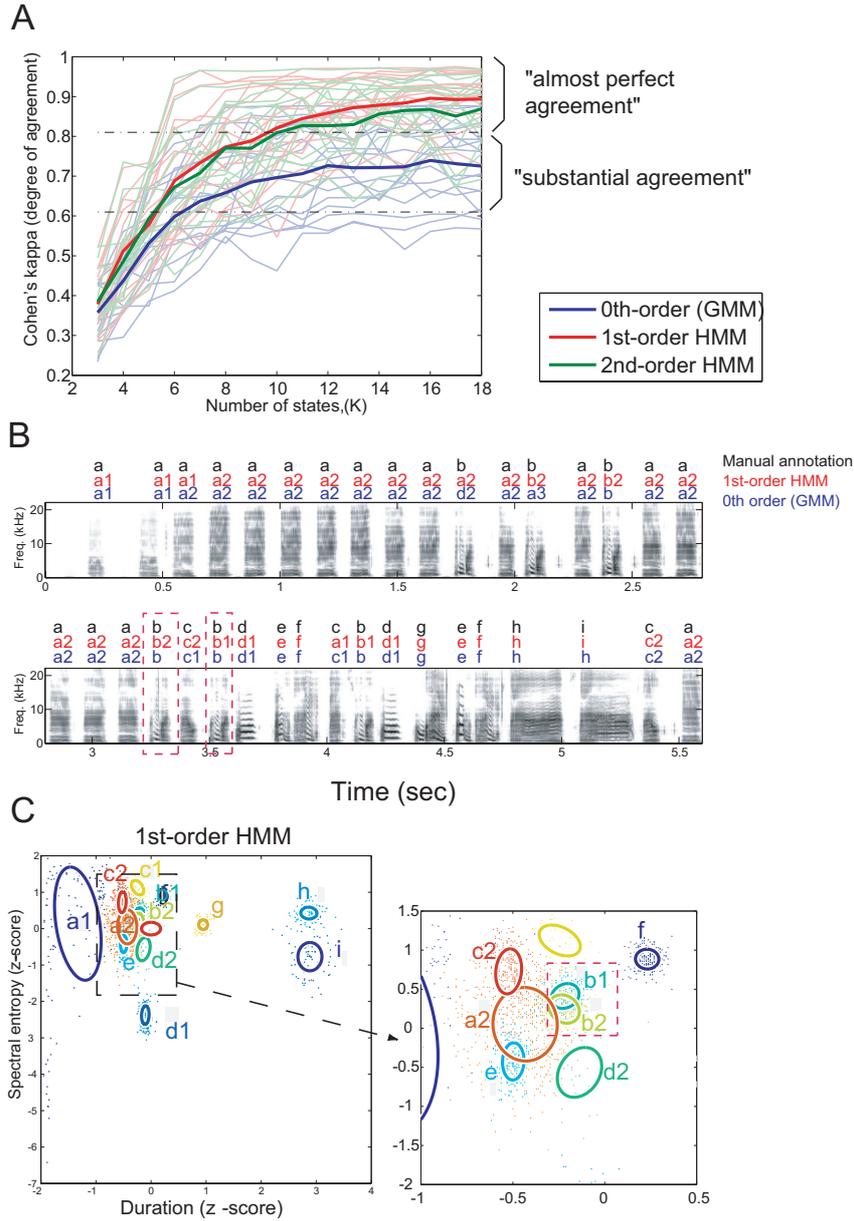}
\end{center}
\caption{Comparison of labeling results using various models and human experts. (A) Kappa's coefficients, which are measure for agreement between model annotations and manual annotation by human experts, are functions of number of states, $K$. Thin lines represent results for individual birds, and thick lines represent average for each model order. (B) Example of annotations for song of BF09. Black labels represent manual annotations done by visual inspection of sonograms. Red and blue labels are labeled using Gaussian mixture model (zeroth-order model) with $K=13$ and first-order HMM with $K=18$, respectively. Number of states $(K)$ that gives the highest lower-bound on log marginal likelihood were used. (C) Example of model fitting results on sound feature space (duration and spectral entropy) for same song as (B). Results from first-order HMM. Ellipses represent contour of Gaussian distribution of each state, and letters indicate syllable aligned to state.} 
\label{fig:annotation_comparison}
\end{figure}

Next we discuss how each model annotates given song syllables. We first compare them with the manual annotations described above (For details in computing the model annotation, see Methods section). 

We evaluated the agreement between the annotations of the models and of human experts by computing Cohen's kappa coefficient, which measures the degree of agreement between two annotators (see Methods section). Figure~\ref{fig:annotation_comparison}A shows the results for all songs. With a sufficient number of hidden states, the average performances (thick lines) of the first- and second-order HMMs reached the region of ``almost perfect agreement'', while the GMM saturated in the region of ``substantial agreement''. Thus, the syllable sequences obtained from the first- and second-order HMMs were in almost perfect agreement with those obtained from manual labeling, while GMM did not provide a comparable result. 

Kappa coefficients for each model-order (selected on the bound) were 0.781 $\pm$ 0.103 (mean $\pm$ s.d.) (range from 0.566-0.905) for the zeroth-order model (GMM), 0.911 $\pm$ 0.055 (range from 0.790-0.978) for the first-order HMM, and 0.873 $\pm$ 0.090 (range from 0.688-0.973) for the second-order HMM. There were significant differences between GMM and the-first order HMM ($p<0.001$), and between GMM and the second-order HMM ($p<0.001$). While the mean performances of the-first order HMM were slightly better than those of the second-order HMM, there was no significant difference between them. The differences between GMM (zeroth-order HMM) and the two HMMs suggest that taking into account the context information improves stable labeling. Human experts may use such context information. 

Next we discuss how each model annotates the syllables. Figure~\ref{fig:annotation_comparison}B compares manual labeling and the labeling of two of models, first-order HMM and GMM, in which the number of hidden states $K$ is selected based on the model selection criteria described in the Methods section. We observed that the two models tended to divide syllables into larger letters. For example, a human labels the first repeated introductory notes using only ``a'', but the two models label them using two states ``a1'' and ``a2''. This may reflect the fact that our method is more sensitive to differences in acoustic features than humans. An important difference between the zeroth-order model and first-order HMM is in the sequence ``bcbd'' by manual labeling. As shown in Figure~\ref{fig:G1_1automaton}, the subsequent syllable after the first ``b'' and the second ``b'' depends on the previous syllable (whether ``a'' or ``c''), and these two syllables "b" are divided into distinct states in the POMM representation (Figure~\ref{fig:G1_1automaton}C). The first-order HMM obtained the following representation: it divided the syllable "b" into the states ``b1'' and ``b2'', while the zeroth-order model, the GMM, did not (red rectangle in Figure~\ref{fig:annotation_comparison}B). This difference is due to context dependency. As we can see in the red rectangle in Figure~\ref{fig:annotation_comparison}C, the distributions of ``b1'' and ``b2'' largely overlap; thus, indistinguishable without using information of the preceding syllables. For the other songs we analyzed, similar properties were often observed: Of the 54 syllables where significant second-order dependency was found in manual annotation-based analysis, the first-order HMM divided 30 syllables into distinct states according to the preceding syllables, while the GMM did so for 17 syllables and the second-order HMM did for 23. As a recent study showed, the contexts affect the acoustic properties~\cite{wohlgemuth2010linked}. Thus, even the GMM, which does not incorporate pre-state dependency, aligned the different states for the same syllables solely on the differences in acoustic features. However, the difference between the GMM and HMMs suggests that HMMs tend to align different syllables depending on the context, not solely on the acoustic features. 

\section{Discussion}
We explored the statistical properties of complex sequencing rules of the Bengalese finch song. To achieve this, we analyzed history dependencies in the syllable sequences annotated by human experts. Then we applied statistical models to acoustic feature data. We discuss the implications of our results and possible neural implementation. 

\paragraph{First-order HMM is sufficient for producing higher-order Markov sequences}
\label{sec:firstHMMisSufficient}

\begin{figure}[tb]
\begin{center}
\includegraphics[width=0.8\linewidth]
{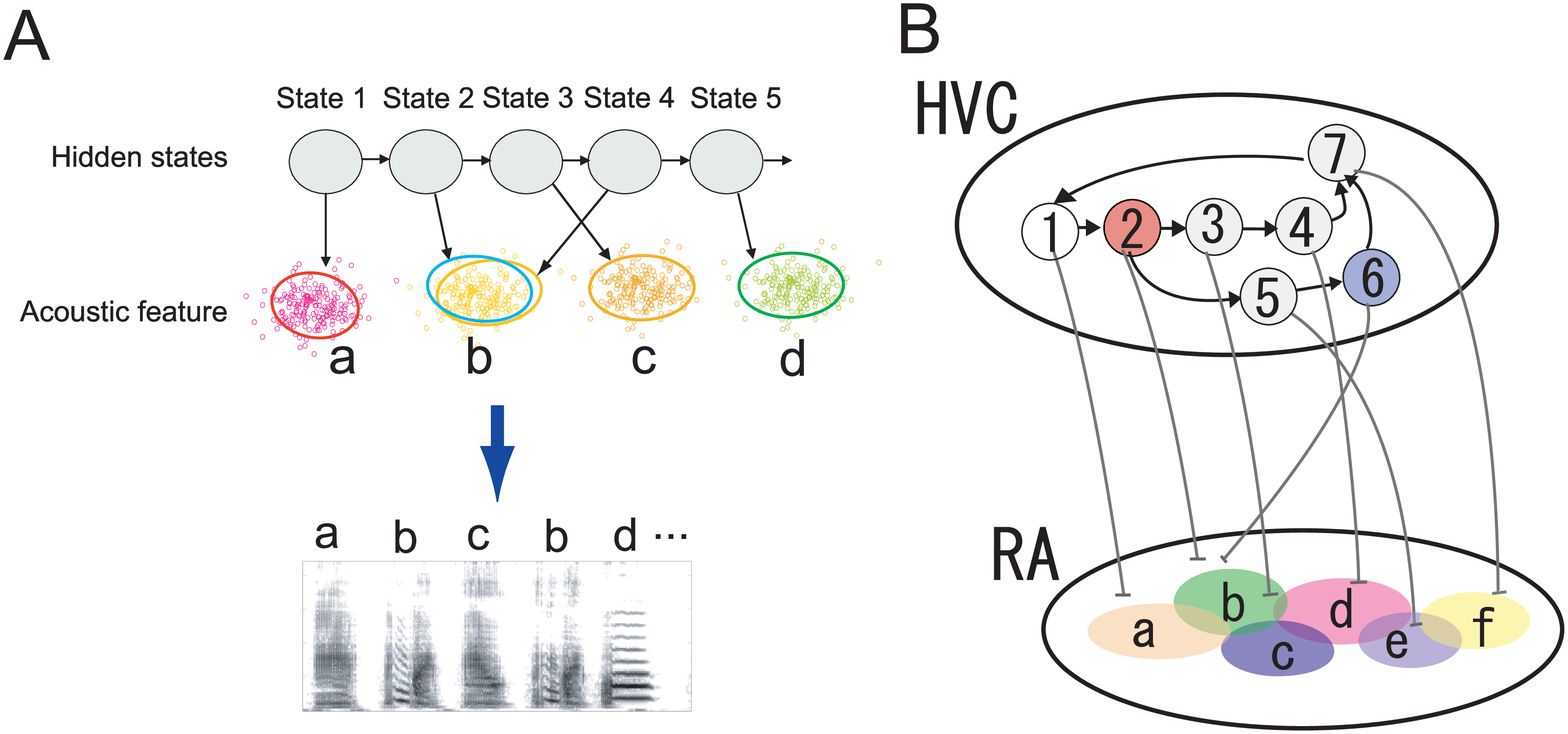} 
\end{center}
\caption{Schematic diagram representing how first-order HMM generates sequences that have higher-order context dependency and its neural implementation. (A) key point is that different states (States 2 and 4) can generate similar acoustic feature space (``b''). This mechanism allows observed song sequences to have higher-order context dependency, even if hidden state sequences are generated from simple Markov process. (B) Schematic of proposed model for neural implementation of Bengalese finch song syntax. Each circle in HVC represents neuron group consisting of feedforward chain of RA-projection neurons. Each group in HVC plays role in generating particular syllables through neuron group in RA. Groups 1 and 6 in HVC project to same RA neuron group that generates song syllable ``b'', as do States 2 and 4 in (A).} 
\label{fig:schematic_1stHMM}
\end{figure}

We have seen that in all songs we examined, the first-order HMMs showed comparable or superior performance (i.e., gave a larger lower bound, better agreement with manual annotation, and cross-validated prediction error) compared with the second-order HMMs when there were enough hidden states, whereas when the number of states was small, the second-order HMMs performed better. These results suggest that (1) when the number of hidden states (labels) is small, considering the higher-order context dependency among hidden states, it leads to a better explanation of the birdsong data, but (2) when we use enough hidden states, the first-order HMMs become sufficient for explaining the data. Such models give better descriptions of data by using a smaller parameter set than higher-order HMMs. 

These findings perhaps are counterintuitive given that our first observation was that some syllables in the Bengalese finch song sequences have dependence on at least two previous syllables. We interpreted our results as follows. Even if the hidden state sequences of the first-order HMMs have only first-order dependency, the emitting syllable acoustic features can have higher-order dependency. This can occur when the different hidden states (States 2 and 4 in Figure~\ref{fig:schematic_1stHMM}A) have similar emission distributions (corresponding syllable ``b'' in Figure~\ref{fig:schematic_1stHMM}A). Although the hidden state sequence is a first-order Markov sequence (``12345...''), the emitted syllables can have second-order dependency (``abcbd...''). This representation (we call POMM) was that the first-order HMM was indeed attained through an automatic parameter fitting process, as we saw in the Results section. 

By adopting this representation, we can avoid exponential growth in the number of parameters when second-order context dependency is sparse. Let us consider an extreme case where all syllables truly depend on two previous syllables (i.e., all components of the transition matrix from the previous two states to the next state are mutually independent). For this case, if we use first-order HMMs to represent the data in the above manner, we need $K^2$ hidden states. This requires a $K^2 \times K^2$ transition matrix, which is larger than that of second-order HMMs ($K^2 \times K$). In addition, first-order HMMs have parameters for $K^2$ Gaussian components, whereas second-order HMMs have those for only $K$ components. If two different models can represent the true data distribution, the model with fewer parameters gives a larger marginal log likelihood and better predictive performance. Thus, in this case, a  second-order HMM will be selected. On the other hand, if most syllables depend only on one last state, and some portion of the syllable depends on the two previous syllables, a first-order HMM can represent the statistical structure of the sequence and will be selected. Our results suggest that the statistical structure of the Bengalese finch song is close to the latter. 

\paragraph{Possible neural implementation} 
These results motivate us to discuss neural implementation of this statistical model structure. In songbirds, two nuclei are mainly related to generating songs: the HVC (proper name) and the robust nucleus of the archistriatum (RA). Neural activity in the HVC appears to encode sequential information~\cite{yu1996temporal,hahnloser2002ultra}, while RA encodes the acoustic structure of individual song syllables~\cite{yu1996temporal,leonardo2005ensemble,sober2008central}. The HVC projects to the RA, while the RA projects to the nuclei that control the syrinx and respiration muscles. The sequential pattern in birdsongs is assumed generated in a feedforward chain of RA-projecting neurons in the HVC~\cite{hahnloser2002ultra,jin2007intrinsic,long2008using}. A theoretical study has shown that such a feedforward-chain mechanism can be extended to generate stochastic branching sequences that obey a first-order Markov process~\cite{jin2009generating}. Whether it can be extended to a higher-order Markov process is unknown. Our results imply that a feedforward chain mechanism that obeys the first Markov process would suffice and be suitable to explain the song syntax of the Bengalese finch. We propose a mechanism that relates our statistical model and the neural circuits. First, we relate a group of feedforward chains to the hidden state of a  first-order HMM. Figure~\ref{fig:schematic_1stHMM}B illustrates this model. The circles in the HVC represent the neuron groups consisting of a feedforward chain of RA-projecting neurons. The arrows represent the firing order of the chain. The firing order is a first-order Markov process that includes stochastic transitions from group 2 to group 3 or group 5. We assume that different groups (groups 1 and 6) generate the same syllables, ``a'', by having a similar projection pattern to the RA. Due to this mechanism, even though the firing order of the HVC(RA) neurons in the HVC is a first-order Markov process, the observable syllable sequences obeys higher-order Markov processes, as observed in our song analysis with the first-order HMM. 

\paragraph{Future Work} 
We showed that the first order HMM gives annotations that were close to manual annotations compared to a standard clustering technique (GMM), which does not use context information. This result suggests that our method with an ordinary HMM can be used as a convenient tool for annotating the Bengalese finch song, instead of time-consuming manual annotation. We applied our method to the songs of only adult healthy Bengalese finches. Investigating the developmental change of the Bengalese finch song or those developed with abnormal conditions such as isolated from song tutors, or with lesions in the song-related nucleus, may be for future study. Such studies will give valuable insight into how complex sequencing rules are formed through learning. 



\section*{Methods}
\paragraph{Recording} 
We analyzed undirected songs (songs in the absence of a female) of 16 adult male Bengalese finches (labeled as BF01-BF16) ranging 133-163 days of age. They were raised in colonies at the RIKEN Brain Science Institute. Before recording, each bird was moved to a sound proof room and isolated from the other birds. Songs were recorded for 24 hours using a microphone placed in the room. All experimental procedures and housing conditions were approved by the Animal Experiments Committee at RIKEN. 

\paragraph{Sound feature extraction} 
To extract acoustic features from each syllable, we used Sound Analysis Pro (SA+) software~\cite{tchernichovski2000procedure}, which is a widely used tool for quantifying song features in birdsongs (~\cite{wu2008statistical} and references therein). We used three representative features: syllable duration, mean pitch, and mean Wiener (spectral) entropy. We applied a feature batch module in SA+ for extracting the acoustic features from wave format audio files. We then randomly picked and analyzed thirty song bouts for each bird from all recordings. For cross-validation, ten additional bouts were also randomly picked. 

\paragraph{Evaluation of agreement of annotations} 
The degree of agreement among three annotators were evaluated using Fleiss's $\kappa$ coefficient~\cite{fleiss1971measuring}, which measures the degree of agreement among more than two annotators. The measure calculates the degree of agreement comparing to that which is expected by chance. For both measures, if $\kappa$-coefficients fall in the range of 0.81-1.00, the result is interpreted as ``Almost perfect agreement''. For the range of 0.61-0.80 - ``Substantial agreement'', 0.41-0.60 - ``Moderate agreement'', 0.21-0.40 - ``Fair agreement'', 0.0-0.20 - ``Slight agreement'', and  $ < 0$ - ``Poor agreement''. 

\paragraph{Evaluation of second-order context dependency} 
To evaluate second-order context dependency, we conducted a hypothesis test for each syllable to verify whether the preceding syllable of the syllable being tested affects the occurrence probability of the next syllable. In particular, we seek the syllable that has both more than one preceding syllable and more than one subsequent syllable. We then tested whether the probabilities of transitions from the syllable depend on the preceding syllable by doing a $\chi^2$ test of goodness of fit between the probability distributions that ignore the preceding syllables and those conditioned on the most frequent preceding syllable~\cite{jin2009generating}. We interpret the syllable having second-order context dependency if $p \le 0.05/n$, where $n$ denotes the number of candidate syllables (the Bonferroni correction). 

\paragraph{Higher-order Hidden Markov Models}

\begin{figure}[tbph]
\begin{center}
\includegraphics[width=0.6\linewidth]
{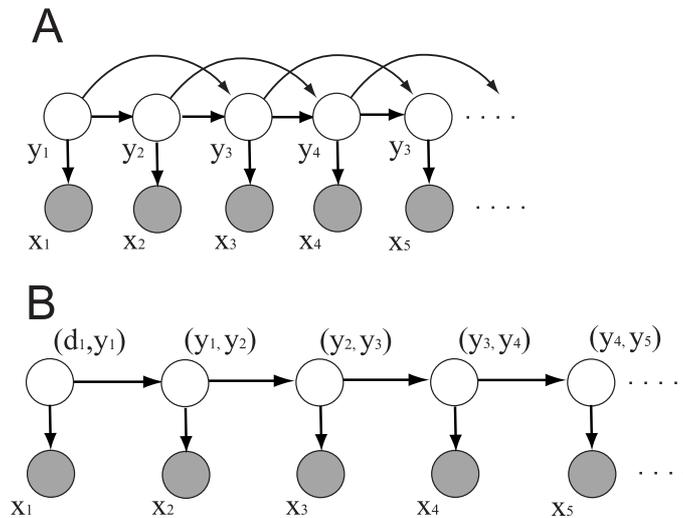} 
\end{center}
\caption{Graphical model representation for second-order HMM describing how parameter estimation for second-order HMM can be done. (A) Naive graphical model for second-order HMM. (B) Another representation of second-order HMM using context states that combine two states. In this graph, we introduce a node (represented as a circle) for each random variable, $x_t, t=1,...,N$. For each conditional distribution, we add arrows to the graph from the nodes corresponding to the variables on which the distribution is conditioned.} 
\label{fig:2ndHMM}
\end{figure}

We consider a second-order HMM, whose directed graphical model is shown in Figure~\ref{fig:2ndHMM}. At first glance, it may seem difficult to apply a forward-backward algorithm, which is the standard algorithm for inferring hidden state sequences, to this model. However, if we combine two succeeding states into one \textit{context states}, we can transform this graphical model into one with the same form as the first-order HMM, as shown in Figure~\ref{fig:2ndHMM}. We introduced a dummy state denoted as $d_1$ for the beginning of the sequences. If we use an $m$-th order HMM, $m-1$ dummy states ($d_1,...,d_{m-1}$) are required. For the second-order HMM, there are $K + K^2$ context states, including ones that contain dummy states. In general, there are $\sum_{l=1}^{m-1} K^{l} + K^m$ context states. For each context state, the number of transition targets is $K$. Hence, the transition matrix $\vec{A}$ has $(\sum_{l=1}^{m-1} K^{l} + K^m) \times K$ elements. 

\paragraph{Annotation analysis} 
To evaluate the agreement between manual annotations by different human annotators and between manual and model annotation, we used Fleiss's kappa coefficient and Cohen's kappa coefficient, respectively. They are statistical measures of inter-annotator agreement for categorical items. They are more robust measures than simple percent agreement calculation since they take into account agreement occurring by chance. Cohen's kappa measures agreement between two raters, while Fleiss' kappa does when there are more than two raters. 

\paragraph{Parameter fitting} 
To train HMMs using given acoustic feature data, we used the Variational Bayes (VB) method~\cite{attias1999inferring,beal2003variational,bishop2006pattern}. The VB method has been widely used as an approximation of the Bayseian method for statistical models that have hidden variables. The VB method approximates true Bayesian posterior distributions with a factorized distribution using an iterative algorithm similar to the expectation maximization (EM) algorithm. For the limit of a large number of samples, the results of the VB coincides with those of the EM algorithm. We used VB because of the following two advantages: (1) its low computational cost, which is comparable to the EM algorithm, and (2) it can select an appropriate model based on the model-selection criterion computed in a model learning process. Full Bayesian approaches based on a sampling technique give a more accurate model-selection criterion, but their high computational cost is unfavorable for our purpose (especially for the second order HMMs, which have a large number of parameters). 

The VB algorithm for the GMM is detailed in \cite{bishop2006pattern}, while those for the first-order HMMs are detailed in~\cite{beal2003variational}. We derived the VB algorithm for the second-order HMMs for the first time, but we only have to change the transition matrix from the algorithm for the first-order HMMs, which is a straightforward extension. If we combine two succeeding states into one context state, we can transform this graphical model into one with the same form as the first-order HMMs, as shown in Figure~\ref{fig:2ndHMM}. We introduced a dummy state denoted as $d_1$ for the beginning of the sequences. If we use an $m$-th order HMM, $m-1$ dummy states ($d_1,...,d_{m-1}$) are required. For the second-order HMMs, there are $K + K^2$ context states, including ones that contain dummy states. In general, there are $\sum_{l=1}^{m-1} K^{l} + K^m$ context states. For each context state, the number of transition targets is $K$. Hence, the transition matrix $\vec{A}$ has $(\sum_{l=1}^{m-1} K^{l} + K^m) \times K$ elements. 

\paragraph{Model selection}
We denote the model index $\idxModel$, which refers to the number of states $K$ and order of Markov process $m$. By using Bayes theorem, the posterior of the model index given data $X^n$ is given by 
\begin{align} 
p(\idxModel|X^n) = \frac{p(X^n|\idxModel) p(\idxModel)}{p(X^n)}. 
\end{align} 

We naturally assume that $p(\idxModel)$ is the uniform distribution, i.e., we have no a priori assumption of the model structure. Then, $p(\idxModel|X^n) \propto p(X^n|\idxModel)$, hence the model gives the highest posterior probability that corresponds to the one that gives the highest marginal log likelihood $p(X^n|\idxModel)$. Ideally, the marginal log likelihood $p(X^n|\idxModel)$ is obtained by marginalization over hidden variable sets (denoted as $Y^n$) and  parameter sets (denoted as $\theta$) as 
\begin{align}
p(X^n|\idxModel) = \sum_{Y^n} \int d \, \vec{\theta} \, p(X^n, 
Y^n|\vec{\theta}, \idxModel) \, p(\vec{\theta}|\idxModel). 
\end{align}
However, this marginalization procedure is infeasible. Therefore, we used a bound on the marginal log likelihood $\log p(X^n|\idxModel)$ instead. The variational free energy ${\cal F}$ gives an upper bound on $- \log p(X^n|\idxModel)$. In other words, the log marginal likelihood $\log p(X^n|\idxModel)$ is lower bounded by negative variational free energy $-{\cal F}$. To emphasize the statistical meanings, we call $-{\cal F}$ the lower bound on the log marginal likelihood. 

\paragraph{Computing model annotation} 
We assumed that each hidden state in the models we used represents one syllable. The models asigned the label that corresponds to the state that gave the highest posterior probability of generating the acoustic features for each syllable (see Methods). The posterior probabilities were computed using the Baum-Welch algorithm~\cite{rabiner1989tutorial}. We then aligned a syllable label to each state so that the aligned labels were the most frequently labeled syllables by human experts in the syllable set that the model state aligned. We allowed more than one state to share the same syllable (many-to-one mapping from states to a syllable). 


\end{document}